\documentclass[10pt,twocolumn,letterpaper]{article}

\usepackage{iccv}
\usepackage{times}
\usepackage{epsfig}
\usepackage{graphicx}
\usepackage{amsmath}
\usepackage{amssymb,pifont}
\usepackage{subcaption}

\usepackage{tabularx}
\usepackage{booktabs}
\usepackage{upgreek}
\usepackage{makecell}
\usepackage{caption} 
\usepackage[table]{xcolor}

\usepackage[pagebackref=true,breaklinks=true,letterpaper=true,colorlinks,bookmarks=false]{hyperref}

\def\iccvPaperID{51} 

\ificcvfinal\fi
\begin{document}
	
	\title{WhiteNNer-Blind Image Denoising via Noise Whiteness Priors}
	
	\maketitle
	\thispagestyle{empty}

\title{WhiteNNer-Blind Image Denoising via Noise Whiteness Priors}
\author{Saeed Izadi}
\author{Zahra Mirikharaji}
\author{Mengliu Zhao}
\author{Ghassan Hamarneh}
\affil{School of Computing Science, Simon Fraser University, Canada\protect \\\texttt{\{saeedi, zmirikha, mengliuz, hamarneh\}@sfu.ca}}

\maketitle
\thispagestyle{empty}
	

\newcommand\figTeaser{
\begin{figure}[!tb]
\begin{center}
\includegraphics[width=8cm]{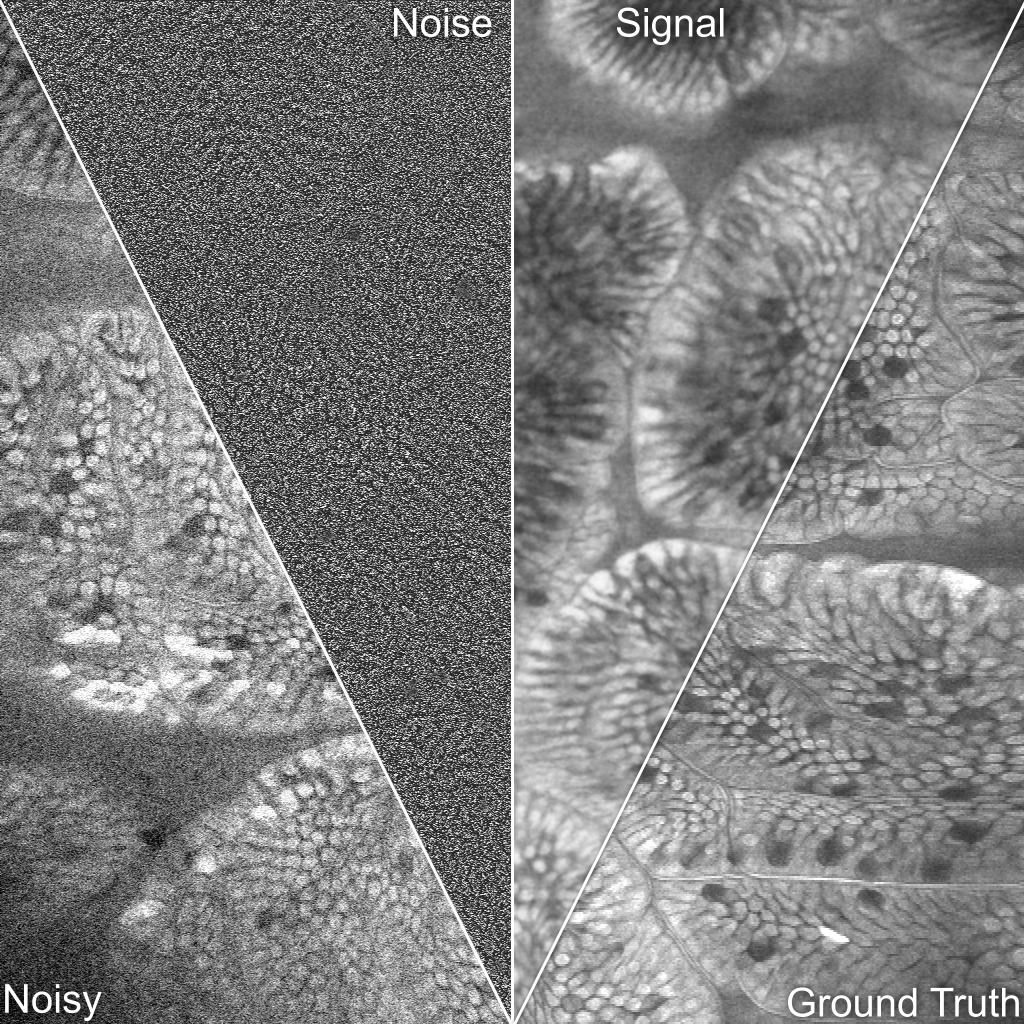}
\end{center}
   \caption{Example of decoupling signal and noise from a single noisy image. \textbf{Signal} and \textbf{Noise} refer to the outputs of the network while \textbf{Noisy} and \textbf{Ground truth} denote the noisy and clean image, respectively. \textbf{Noise} estimate is enhanced via contrast stretching for clarity. 
   }
\label{fig:feat_prior}
\end{figure}
}

\newcommand\figArch{
\begin{figure}[!tb]
\begin{center}
\includegraphics[width=8cm]{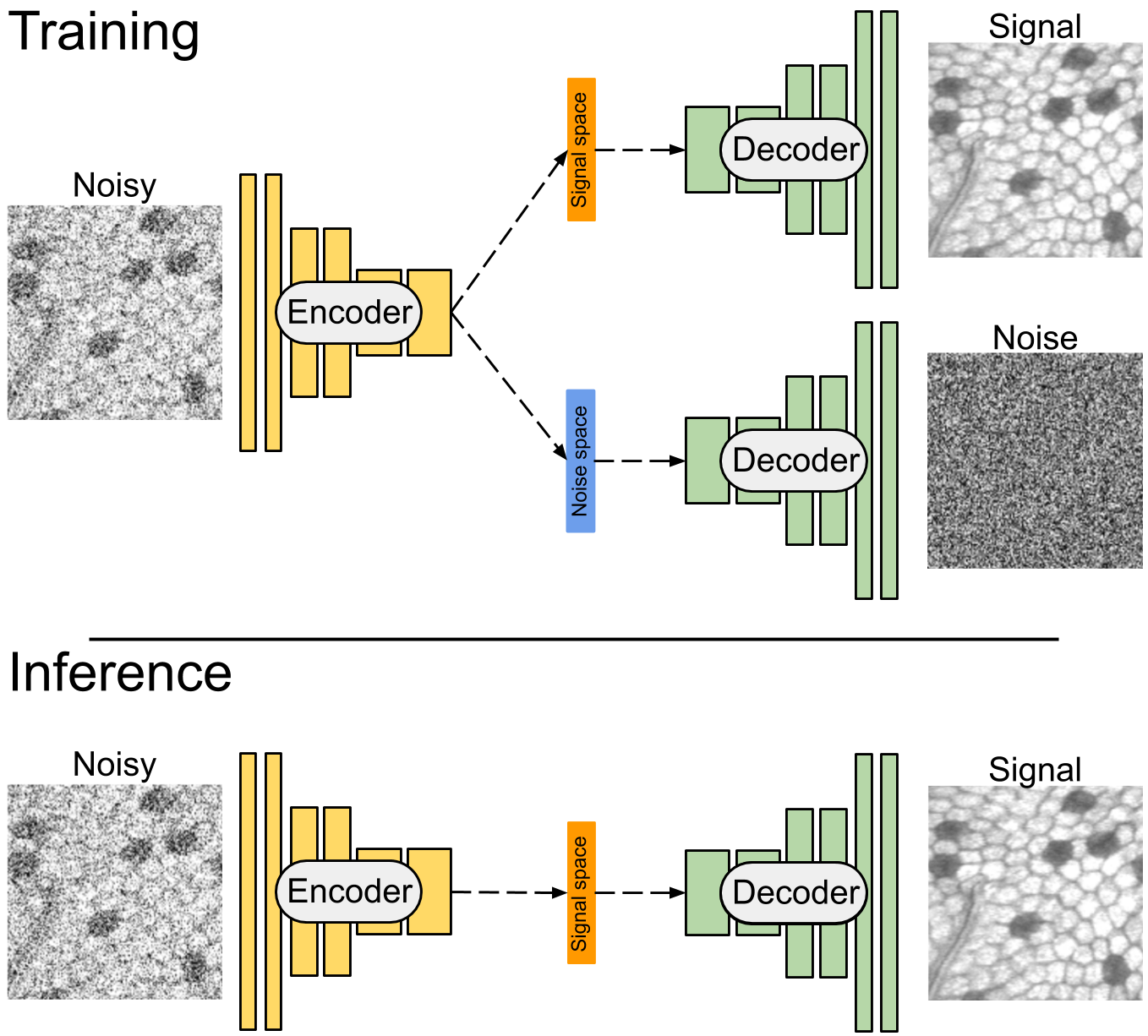}
\end{center}
   \caption{Cap
   }
\label{fig:feat_prior}
\end{figure}
}

\newcommand\figArchh{
\begin{figure*}[!tb]
\begin{center}
\includegraphics[width=\textwidth]{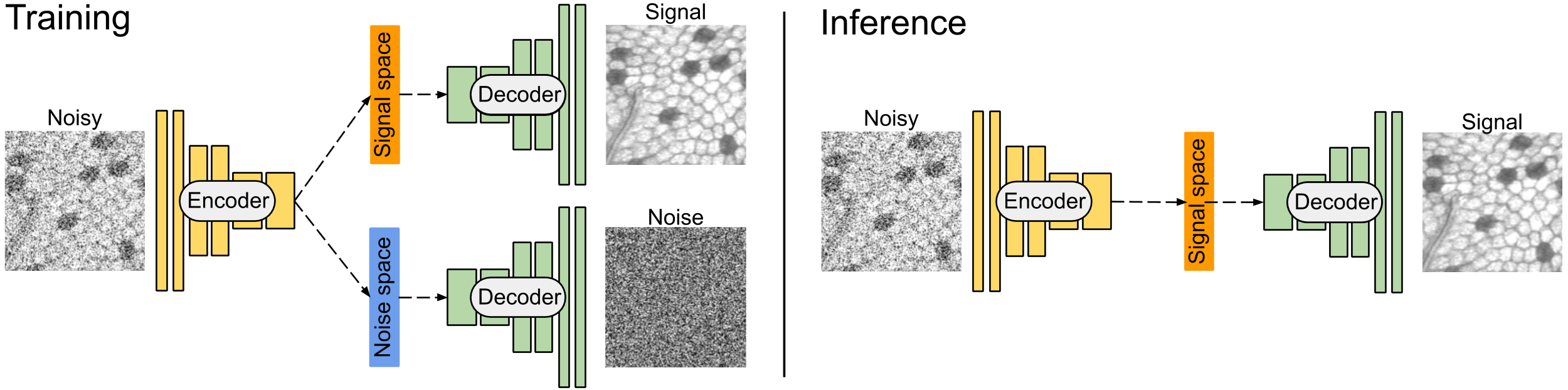}
\end{center}
\caption{Overview of the WhiteNNer network. In training, input noisy image is first processed by an encoder network to decouple the signal and noise in a latent space. Then, both latent representations are fed into a shared decoder to generate the signal and noise in spatial domain. In inference, the latent representation associated with the noise is discarded to only estimate the signal.}
\label{fig:archh}
\end{figure*}
}

\newcommand\figQualitative{
\begin{figure*}[!tb]
\begin{center}
\includegraphics[width=\textwidth]{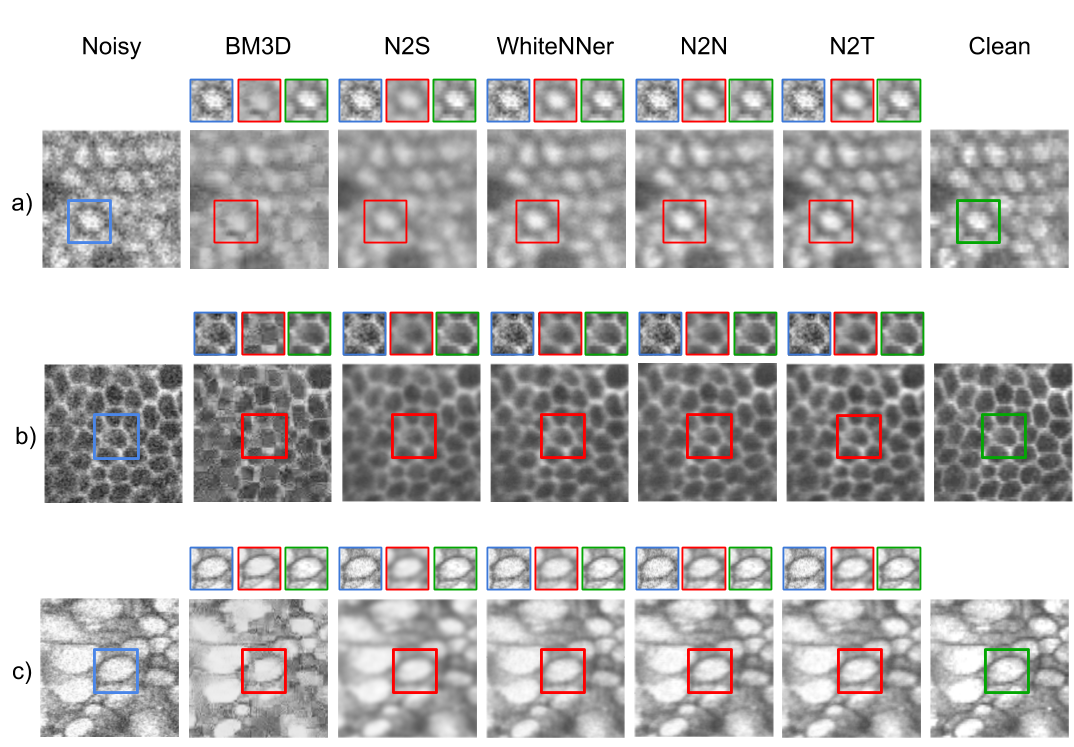}
\end{center}
\vspace{-5mm}
\caption{Qualitative denoising results on three datasets at $\sigma=25$. Each row depicts the evaluation on a) CLE100, b) CLE200, and c) CLE1000 samples. For each method, side-by-side comparison between the obtained denoised image, ground truth and input noisy image is provided in red, green and blue boxes.}
\label{fig:qual}
\end{figure*}
}

\newcommand\figPiecewise{
\begin{figure}[!tb]
\begin{center}
\includegraphics[width=8cm]{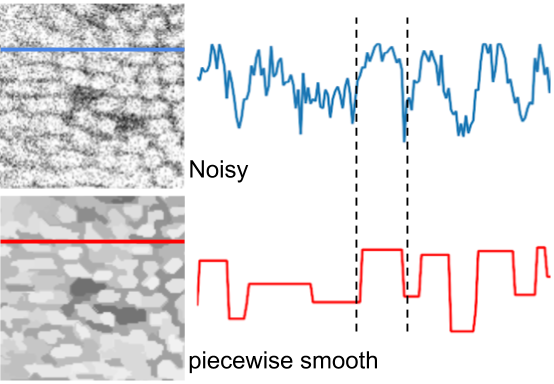}
\end{center}
   \caption{ Example of generating piecewise constant image for single noisy images. Intensity profiles (right) are drawn for a horizontal line across the images (left). 
   }
\label{fig:feat_prior}
\end{figure}
}

\newcommand\figNoiseLevel{
\begin{figure}[!tb]
\begin{center}
\includegraphics[width=8cm]{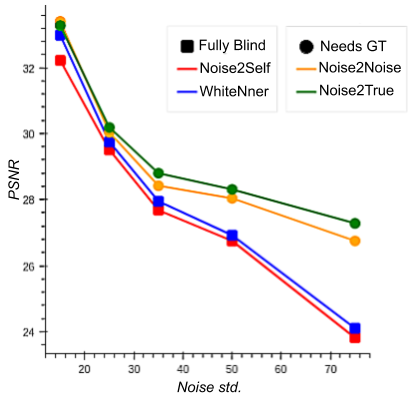}
\end{center}
   \caption{Performance of WhiteNNer against baselines on CLE100 dataset at different noise levels. We use average PSNR scores for the comparison. The square symbol indicates methods that are fully blind while the circle symbol indicates methods that benefit from ground truth for training or generating additional instances of the noisy input. 
   }
\label{fig:noiselevel}
\end{figure}
}

\newcommand{\cmark}{\ding{51}} 
\newcommand{\xmark}{\ding{53}}

\newcommand\tblResults{
\begin{table*}[t!]
\vspace{-3mm}
\caption{Quantitative results of WhiteNNer against baseline methods in term of PSNR and SSIM on held-out test data from three CLE datasets. Columns denoted by $\sigma$, {\footnotesize$ \mathcal{S}$} and {\footnotesize$ \mathcal{N}'$} indicate whether methods leverage noise variance, ground truth, or different realization of the noisy image. Error bars for CNNs is calculated from training five models.
\label{tab:results}
}
\begin{center}
\begin{tabular}{|l|l|l|l||c|c|c|c|c|c|c|c|}
\Xhline{1pt}
\rowcolor[rgb]{ .851,  .851,  .851}
\multicolumn{4}{|c||}{Dataset}
&
\multicolumn{2}{c}{CLE100}
&
\multicolumn{2}{c}{CLE200}
&
\multicolumn{2}{c|}{CLE1000}
\\
\rowcolor[rgb]{ .851,  .851,  .851}
\multicolumn{1}{|c|}{Metric}
&$\sigma$
&\footnotesize $\mathcal{S}$
&\footnotesize $\mathcal{N}'$
&PSNR$\uparrow$&SSIM$\uparrow$
&PSNR$\uparrow$&SSIM$\uparrow$
&PSNR$\uparrow$&SSIM$\uparrow$
\\
\hline
\hline
NLM\ \ ~\cite{Buades2008}
&\cmark&\xmark&\xmark&23.49&0.3236&22.87&0.3318&23.40&0.3383 
\\
BM3D\ \ ~\cite{bm3d2007}
&\cmark&\xmark&\xmark&25.91&0.4152&24.74&0.4297&25.58&0.3877
\\
N2S\ \ ~\cite{noise2self}
&\xmark&\xmark&\xmark&26.75$\pm$0.533&0.4467&25.12$\pm$0.521&0.4817&25.99$\pm$0.450&0.3989
\\
N2N\ \ ~\cite{noise2noise}
&\xmark&\xmark&\cmark&28.04$\pm$0.396&0.5070&26.00$\pm$0.297&0.5699&27.97$\pm$0.149&0.5139
\\
N2T
&\xmark&\cmark&\xmark&28.31$\pm$0.268&0.5128&26.79$\pm$0.495&0.5790&28.29$\pm$0.362&0.5315
\\
\hline
\hline
WhiteNNer-1
&\xmark&\xmark&\xmark&25.75$\pm$0.275&0.4171&24.62$\pm$0.127&0.4627&25.63$\pm$0.401&0.3934
\\
WhiteNNer-2
&\xmark&\xmark&\xmark&27.05$\pm$0.184&0.4895&25.79$\pm$0.275&0.5301&26.58$\pm$0.245&0.4694

\\
\hline
\end{tabular}
\vspace{-7mm}
\end{center}

\label{t2}
\end{table*}
}

\newcommand\tblNoiseLevel{
\begin{table}[t!]
\vspace{-3mm}
\caption{Quantitative results of WhiteNNer against baseline methods for different noise levels. 
\label{tab:noiselevel}
}
\begin{center}
\resizebox{\columnwidth}{!}{%
\begin{tabular}{|l||c|c|c|c|}
\Xhline{1pt}
\rowcolor[rgb]{ .851,  .851,  .851}
\multicolumn{1}{|c||}{Dataset}
&$\sigma=25$
&$\sigma=35$
&$\sigma=50$
&$\sigma=75$

\\
\rowcolor[rgb]{ .851,  .851,  .851}
\multicolumn{1}{|c||}{Metric}
&PSNR$\uparrow$
&PSNR$\uparrow$
&PSNR$\uparrow$
&PSNR$\uparrow$
\\
\hline
\hline
Noise2Self\ \ ~\cite{noise2self}
&29.51&27.68&26.75&23.81 
\\
Noise2Noise\ \ ~\cite{noise2noise}
&30.04&28.42&28.04&26.75 
\\
Noise2True
&30.19&28.80&28.31&27.28 
\\
\hline
\hline
WhiteNNer-1
&29.49&27.28&25.75&22.73 
\\
WhiteNNer-2
&29.67&27.85&26.84&23.91
\\
\hline
\end{tabular}
}
\vspace{-7mm}
\end{center}
\end{table}
}

\newcommand\tblData{
\begin{table*}[t!]
\caption{Summary of datasets used in our experiments.}
\label{tab:cledatasets}
\begin{center}
\begin{tabular}{|l|l|c|c|c|c|}
\Xhline{1pt}
\rowcolor[rgb]{ .851,  .851,  .851}
Dataset&provided by&\#patients &\#images&anatomical site&image size
\\
\hline
\hline
CLE100&Leong et al.~\cite{cle100}&30&181&small intestine&$1024 \times 1024$\\
         \hline
         CLE200&Grisan et al.~\cite{cle200}&32&262&esophagus&$1024 \times 1024$\\
         \hline
         CLE1000&{\c{S}}tef{\u{a}}nescu et al.~\cite{cle1000}&11&1025&colorectal mucosa&$1024 \times 1024$\\
         \hline
\end{tabular}
\end{center}
\end{table*}
}

\newcommand\tblStatSig{
\begin{table}[t!]
\vspace{-3mm}
\caption{Paired t-test on the CLE100 dataset at $\sigma=50$.
\label{tab:stats}
}
\begin{center}
\resizebox{\columnwidth}{!}{%
\tiny
\begin{tabular}{|l||c|c|}
\Xhline{1pt}
\rowcolor[rgb]{ .851,  .851,  .851}
\multicolumn{1}{|l||}{Method}
&$p$-value
&$t$-value

\\
\hline
\hline
NLM\ \ ~\cite{Buades2008}
&1.35e\textsuperscript{-16}&12.83
\\
BM3D\ \ ~\cite{bm3d2007}
&1.25e\textsuperscript{-12}&9.44
\\
Noise2Self\ \ ~\cite{noise2self}
&2.69e\textsuperscript{-3}&3.11

\\
\hline
\end{tabular}
}
\vspace{-7mm}
\end{center}
\end{table}
}

\newcommand\tblAblation{
\begin{table}[t!]
\vspace{3mm}
\caption{ Performance of different loss function in terms of average PSNR. The results are reported on the CLE100 dataset at $\sigma=50$
\label{tab:ablation}
}
\begin{center}
\resizebox{\columnwidth}{!}{%
\scriptsize
\begin{tabular}{|l||c|}
\Xhline{1pt}
\rowcolor[rgb]{ .851,  .851,  .851}
\multicolumn{1}{|c||}{Loss}
&PSNR$\uparrow$

\\
\hline
\hline
$\mathcal{L}_{rec}+\mathcal{L}_{pc}$\ \ 
&22.41
\\
$\mathcal{L}_{rec}+\mathcal{L}_{pc}+\mathcal{L}_{tv}$\ \ 
&23.14
\\
$\mathcal{L}_{rec}+\mathcal{L}_{pc}+\mathcal{L}_{tv}+\mathcal{L}_{ar}$\ \ 
&25.88
\\
$\mathcal{L}_{rec}+\mathcal{L}_{pc}+\mathcal{L}_{tv}+\mathcal{L}_{st}$\ \ 
&26.02
\\
$\mathcal{L}_{rec}+\mathcal{L}_{pc}+\mathcal{L}_{tv}+\mathcal{L}_{ar}+\mathcal{L}_{st}$\ \ 
&26.12

\\
\hline
\end{tabular}
}
\vspace{-7mm}
\end{center}
\end{table}
}

\begin{abstract}
   The accuracy of medical imaging-based diagnostics is directly impacted by the quality of the collected images. A passive approach to improve image quality is one that lags behind improvements in imaging hardware, awaiting better sensor technology of acquisition devices. An alternative, active strategy is to utilize prior knowledge of the imaging system to directly post-process and improve the acquired images. Traditionally, priors about the image properties are taken into account to restrict the solution space. However, few techniques exploit the prior about the noise properties. In this paper, we propose a neural network-based model for disentangling the signal and noise components of an input noisy image, without the need for any ground truth training data. We design a unified loss function that encodes priors about signal as well as noise estimate in the form of regularization terms. Specifically, by using total variation and piecewise constancy priors along with noise whiteness priors such as auto-correlation and stationary losses, our network learns to decouple an input noisy image into the underlying signal and noise components. We compare our proposed method to Noise2Noise and Noise2Self, as well as non-local mean and BM3D, on three public confocal laser endomicroscopy datasets. Experimental results demonstrate the superiority of our network compared to state-of-the-art in terms of PSNR and SSIM.  
\end{abstract}

\section{Introduction}
Globally, colorectal cancer is the third and the second most commonly diagnosed cancer in men and women, respectively. In 2018, 575,789 men and 520,812 women had a history of colorectal cancer, resulting in an estimated 551,269 deaths~\cite{bray2018global}. Timely inspection of suspicious areas within the gut, followed by a precise diagnosis, is critical for improved disease prognosis and reduced mortality. However, conventional tissue sectioning and \textit{ex-vivo} histological examination are associated with invasive biopsy collection and preparation that significantly delay the screening procedure.  \par

\figTeaser
\figArchh

Handheld, portable confocal laser endomicroscopy (CLE) is a well-established imaging technique that provides a real-time, \textit{in-vivo}, and biopsy-free histological assessment (so-called optical biopsy) of the mucosal layer of the gastrointestinal (GI) tract based on both endoscopic and endomicroscopic images acquired during an ongoing endoscopy. CLE provides magnified visualization of tissues in cellular and subcellular resolution and enables the endoscopists to assess the pathology in gastrointestinal tissue sites, e.g., Barrett’s esophagus~\cite{KIESSLICH2006979} or colonic mucosa~\cite{KIESSLICH2004706}. \par

As the accuracy of medical imaging-based diagnostics depends on the quality of the images, it is critical to curtail imaging noise. One approach for noise reduction is to rely on higher quality image acquisition, be that through longer acquisition times, more elaborate optics, high fidelity electronics, or more complex image reconstruction algorithms. However, all these strategies are not without their drawbacks, i.e., subjecting the patient to lengthier procedures, less portable and more invasive imaging devices, increased cost, or lower frame rates, respectively.  An alternative strategy to enhance the image quality is to directly post-process the corrupted images. \par

Due to the ill-posed nature of the image denoising problem, a large body of classic methods utilizes different regularization techniques, like incorporating additional domain-specific prior knowledge in the model to restrict the possible candidates in the optimization search space by penalizing solutions deemed undesirable~\cite{RUDIN1992259, bm3d2007, ncsr20013, Buades2008}. Despite the advantage of leveraging priors about the properties of the true signal, exploiting available information about the noise is often completely ignored. Conversely, the estimate of the clean image can be further improved once we also ensure that the noise estimate conforms to the known and expected properties of the true imaging noise as well.\par

Recently, convolutional neural networks (CNNs) have shown remarkable performance for image restoration, particularly image denoising~\cite{dncnn2017, ffdnet2018, memnet2017}. These CNN-based approaches are typically trained in a supervised manner, requiring pairs of noisy and clean images (ground truth). However, in many applications, it is practically impossible to acquire the ground truth images, e.g., medical imaging of real patients. Lehtinen et al.~\cite{noise2noise} proposed the Noise2Noise training scheme to relax the requirement of providing ground truth by allowing the network to learn the mapping between two instances of the noisy image containing the same signal. Despite the efforts to remove the need for clean images in the training procedure, it is unfortunately still impractical to acquire several instances of the noisy image in some contexts, particularly for endomicroscopy. Alternatively, Batson et al.~\cite{noise2self} proposed Noise2Self which is a self-supervised approach to denoise a corrupted image from only a single noisy instance. \par

In this paper, in addition to relying on explicit signal priors, we propose incorporating novel regularization terms into the CNN loss function thus encouraging the noise to respect the whiteness priors. Using these priors, we train our network to map the noisy images to the clean and noise components, without any ground truth. Moreover, instead of computing the noise by subtracting the predicted signal from the noisy input, we employ a neural architecture with two tails -- one for inferring the clean image and the other for the noise. This design can serve as a new architectural regularization to facilitate the separability of signal and noise in a latent space, leading to the better reconstruction of both. 

In a nutshell, the main contributions of the paper are summarized as follows:
\begin{description}
\setlength{\itemindent}{0.0in}
  \item[$\bullet$] We propose a novel blind denoising model that leverages both signal and noise priors in the form of regularization terms in a CNN loss function.
  \item[$\bullet$] We present the first network that is capable of decoupling signal and noise without any ground truth.
  \item[$\bullet$] We demonstrate that our proposed method outperforms state-of-the-art blind denoising techniques on three confocal endomicroscopy datasets. 
\end{description}

\section{Related Works}
Most existing image denoising techniques in the literature leverage prior knowledge and we categorize them into the following three general groups: \par
\textbf{Data-driven priors.} Instead of explicit consideration, the approaches in this group encode the priors implicitly through learning the direct mapping between pairs of corrupted and clean images. Thanks to the strength of deep networks, recent denoising methods have achieved impressive results. The first attempt to leverage neural networks for image denoising was conducted by Burger et al.~\cite{mlp2012} who used a simple multi-layer perceptron architecture for the learning. Afterward, Zhang et al.~\cite{dncnn2017} demonstrated that residual learning and batch normalization not only improve the denoising performance but also expedite the training procedure. Thai et al.~\cite{memnet2017} proposed a very deep memory persistent network by stacking dozens of memory units that are densely connected. Zhang et al.~\cite{ffdnet2018} suggested feeding a noise level map as well as the noisy image to the network. Doing so, the network will be able to handle different levels of corruption and spatially-variant noise. Most recently, the network proposed by Guo et al.~\cite{Guo_2019_CVPR} consists of one noise level estimator and one non-blind denoiser sub-network to robustly generalize the model performance on real-world noise. \par

\textbf{Signal Priors.} Approaches in this group, which encompasses the majority of the traditional image denoising techniques, exploit the hand-crafted priors about the underlying signal, such as smoothness~\cite{buades2005review}, gradient~\cite{RUDIN1992259}, sparsity~\cite{sparsity2006} and non-local self-similarity~\cite{bm3d2007}. Classic approaches for image denoising can be traced back to the piecewise constant prior in the Mumford-Shah model, which was originally proposed for image segmentation~\cite{mumford1989optimal}, and then extended by Tsai tet al.~\cite{Tsai2001} for image denoising. Later, Rudin et al.~\cite{RUDIN1992259} exploited total variation (TV) prior in a variational formulation to preserve sharp edges of the underlying signal while suppressing the noise. Non-local self-similarity (NSS) prior relates to the fact that images often contain many similar yet non-local (i.e. spatially distant) patterns within the image. BM3D~\cite{bm3d2007} is an NNS-based denoising method which groups similar patches, transforms the groups into the frequency domain and attenuates the noise by hard-thresholding of the transform coefficients. More recently, Ulyanov et al.~\cite{Lempitsky2018} proposed to leverage the deep networks with randomly-initialized weights as a hand-crafted prior for image denoising. In their method, the network is trained to reconstruct the corrupted image from random noise input with an early stop constraint to prevent fitting to the noise. One can interpret their work as an NNS-based algorithm since a linear combination of spatially shared kernels is used to generate a clean image. \par

\textbf{Noise Priors.} There have been several attempts to exploit prior knowledge about spectral whiteness of the residual images in deblurring problem~\cite{Almeida2013, Hansen2006, rust2008residual}. However, to the best of our knowledge, only a few have focused on image denoising. Lanza et al.~\cite{lanza2013variational} proposed to explicitly enforce whiteness property by imposing soft constraints on the auto-correlation of the residual image in the frequency domain. Recently, Soltanayev ~\cite{NIPS2018_7587} leveraged the Gaussianity property of the AWGN to adopt Stein’s Unbiased Risk Estimator (SURE) as a loss function for training the deep networks without clean ground truth images. 

\section{Method}
Image denoising refers to the process of inspecting the noisy image to decouple the underlying signal from the noise component. One common assumption is that the noise is additive white Gaussian (AWGN) with a standard deviation of $\sigma$. Let us consider $\mathcal{X} = \{x_{i} \in \mathbb{R}\}_{i=1}^{M}$ to represent the noisy image formed by adding random white Gaussian noise $\mathcal{N} = \{n_{i} \in \mathbb{R}\}_{i=1}^{M}$ to a clean signal $\mathcal{S} = \{s_{i} \in \mathbb{R}\}_{i=1}^{M}$ where $M$ denotes total number of pixels. Letting $i$ to denote the index for a single pixel, the degradation model can be written as follows:
\begin{equation}
    x_i = s_i + n_i
\end{equation}
\begin{equation*}
    n_i  \backsim \mathrm{N}(0, \sigma)
\end{equation*}
The goal of our model is to take a single noisy image and decouple it back into the signal $\mathcal{S}$ and noise $\mathcal{N}$, without any ground truth. To do so, we need to design accurate and discriminative priors for either component and use them as the supervisory signal during the training.

\figPiecewise

\subsection{Architecture and Inference}
As depicted in Fig~\ref{fig:archh}, our architecture embodies the encoder-decoder paradigm with skip connections~\cite{UNet} (removed in Fig~\ref{fig:archh} for simplicity). The encoder $\phi_{enc}$ takes the noisy image $\mathcal{X}$ where each pixel is normalized to $[0,1]$ and outputs two $p$-dimensional latent features $\mathcal{F}^{\mathcal{N}} = \{f_{i}^{n} \in \mathbb{R}\}_{i=1}^{p}$, $\mathcal{F}^{\mathcal{S}} = \{f^{s}_{i} \in \mathbb{R}\}_{i=1}^{p}$, The features $\mathcal{F}^{\mathcal{S}}$ and $\mathcal{F}^{\mathcal{N}}$ are then decoded to the spatial domain by a shared decoder $\phi_{dec}$, resulting in signal $\mathcal{S}$ and noise $\mathcal{N}$ estimates. Accordingly, our network has a single input and two outputs. Once the model is trained, the latent representation for the noise $\mathcal{F}^{\mathcal{N}}$ can be simply discarded as it is only provided to serve as an architectural regularizer. In other words, the models maps the $\mathcal{X}$ to both $\mathcal{F}^{\mathcal{S}}$ and $\mathcal{F}^{\mathcal{N}}$ while only $\mathcal{F}^{\mathcal{S}}$ is fed into the decoder to reconstruct signal $\mathcal{S}$. \par 
We use the original U-Net~\cite{UNet} architecture for all our experiments with depth 5, kernel size 3, bilinear upsampling and linear activation in the last layer. \par

\subsection{Proposed Loss Functions}
Our model adopts a multi-loss objective function which encodes reconstruction, a.k.a data fidelity, and priors about the properties of the signal and noise. The proposed loss function can be formulated as follows:
\begin{equation}
    \mathcal{L}_{total}(\mathcal{X}, \mathcal{S}, \mathcal{N}) = \mathcal{L}_{rec}(\mathcal{X},\mathcal{S}, \mathcal{N}) + \mathcal{L}_{noise}(\mathcal{N}) +  \mathcal{L}_{signal}(\mathcal{S})
\end{equation}
The noise prior $\mathcal{L}_{noise}$ consists of two distinct terms. The first term, auto-correlation loss $\mathcal{L}_{ac}$, leverages the statistical fact that a white noise image contains pixels intensities which are spatially uncorrelated, given the signal. The second term, stationary loss $\mathcal{L}_{st}$, penalizes the network if the variance of the noise estimate is not consistent spatially. Hence, the loss term for noise priors can be written as: 
\begin{equation}
    \mathcal{L}_{noise}(\mathcal{N}) = \mathcal{L}_{ac}(\mathcal{N}) + \mathcal{L}_{st}(\mathcal{N})
\end{equation}
Turning to signal properties, we take two well-established priors into consideration: the piecewise constancy~\cite{mumford1989optimal} and minimal total variation~\cite{RUDIN1992259} denoted by $\mathcal{L}_{pc}$ and $\mathcal{L}_{tv}$, respectively. Therefore, signal prior can be written as follows: 
\begin{equation}
    \mathcal{L}_{signal}(\mathcal{S}) = \mathcal{L}_{pc}(\mathcal{S}) + \mathcal{L}_{tv}(\mathcal{S})
\end{equation}
\noindent \textbf{Reconstruction Loss.} Given the signal and noise estimates $\mathcal{S}$ and $\mathcal{N}$ in the output, $\mathcal{L}_{rec}$ guarantees that the addition of these two components perfectly reconstructs the original input. We use $L_2$ distance to measure the faithfulness over all pixels. Mathematically, 
\begin{equation}
    \mathcal{L}_{rec}(\mathcal{X},\mathcal{S},\mathcal{N} ) = \frac{1}{M} \sum_{i=1}^{M} [ x_i - (s_i + n_i) ]^2
\end{equation}

\noindent \textbf{Auto-correlation Loss.} In statistics, auto-correlation (AC) is the metric to measure the correlation or similarity between a random process and a time-lagged version of itself. For a discrete random process $y$, the AC at lag $\tau$ is expressed as:
\begin{equation}
    \mathcal{R}(\tau) = E[y(t)\cdot y(t-\tau)]
\end{equation}
In context of 2D images, the lag can be considered in spatial domain across horizontal and vertical directions. Let $I(i,j)$ denote an image in 2D space where $i$ and $j$ are the pixel coordinates. We can define the AC function as a mapping from pixel coordinates to a scalar value, formulated as follows:
\begin{equation}
    \mathcal{R}(\ell, m) = E[I(i, j) I(i-\ell, j-m)]
\end{equation}
where $\ell$ and $m$ are the spatial lag between two distinct pixel coordinates across the horizontal and vertical axis, respectively. On the other hand, the whiteness property of the noise implies every pixel to be uncorrelated to any other pixel. Alternatively stated, given a white noise image, the AC at lag $\ell=0$ and $m=0$ equals to the noise variance $\sigma^2$ while being zero elsewhere, i.e.:
\begin{equation}
    \mathcal{R}(\ell, m)  =\left\{\begin{array}{ll}{\sigma^2} & {\text { if }(l, m)=(0,0)} \\ {0} & {\text { otherwise }}\end{array}\right.
\end{equation}
Moreover, we assume that the noise is ergodic and the image to be denoised is sufficiently large. Following our notation, sample auto-correlation can be used to estimate the AC for a noise image $\mathcal{N}$, defined as: 
\begin{equation}
    \bar R(\ell)  = \frac{1}{M}\sum^{M}_{i=1} n_i \cdot n_{i+\ell}
\end{equation}
To implement $\mathcal{L}_{ac}$,  we use sample AC and minimize it for any lag value greater than zero to penalize noise estimates that are spatially correlated. Precisely, we first pad each side of the noise prediction with a reflected copy of itself. Then, the sample AC is calculated by selecting a random lag in the range of $\ell \in [1, M]$ for each update. \par

\tblData

\noindent \textbf{Stationary Loss.} In general, a random process is said to be stationary if its statistics are not changed over time. White noise image is a simple example of a stationary process where the variance and mean are invariant to the spatial translation. Particularly for the variance statistic, we can mathematically define the stationary property of a noise $\mathcal{N}$ as below:
\begin{equation}
    Var[n_i] = Var[n_{i+\ell}]
\end{equation}
where $\ell$ denotes the translation shifts. To compel the satisfaction of stationary property, we first partition the predicted noise into $B$ non-overlapping $b \times b$ blocks $\{\mathcal{B}_i\}_{i=1}^{B}$  followed by computing the standard deviation $\hat \sigma_b$ within each block $\mathcal{B}_i$, resulting in a set $\mathcal{V} =\{\hat \sigma_b \in \mathbb{R}\}_{b=1}^{B}$. Being a stationary noise implies all elements of $\mathcal{V}$ to be the identical, i.e.:
\begin{equation}
    \hat \sigma_i = \hat \sigma_j ; \forall i,j \in \{1,2,...,B \}
\end{equation}
Therefore, we apply a Softmax on the elements in $\mathcal{V}$ to get a probability distribution $\Uppsi = \{\psi_i \in [0,1]\}$ over all blocks. Ideally, $\Uppsi$ should give the same probability estimation for every block. To measure this, we compute $\mathcal{L}_{st}$ by computing the cross-entropy between $\Uppsi$ and a discrete uniform distribution over $B$ blocks. In our implementation, we randomly select $b$ from $\{2,4, 8, 16\}$ for each update.\par

\noindent \textbf{Piecewise Constant Loss.} Based on the prior suggested by Mumford Shah~\cite{mumford1989optimal}, $\mathcal{L}_{pc}$ is designed to encourage our model to output signal estimates that contain constant intensity values for all pixels within very small segments. To compute $\mathcal{L}_{pc}$, we first need to simulate the piecewise constant counterpart of the noisy input. To do so, a graph-based segmentation~\cite{Felzenszwalb2004} method is firstly used to partition the noisy image $\mathcal{N}$ into $K$ pixel clusters $\{\mathcal{C}_k\}_{k=1}^{K}$, where $\mathcal{C}_k$ refers to the set of indices of pixels that belong to the $k$th cluster. We subsequently replace the values of pixels within each cluster with their average intensities, resulting in $\mathcal{M} = \{m_i \in \mathbb{R}\}_{i=1}^{M}$. Afterward, we measure $\mathcal{L}_{pc}$ by computing the $L_{2}$ distance between the gradients of the signal estimate $\mathcal{S}$ and simulated piecewise constant image $\mathcal{M}$: 
\begin{equation}
    \mathcal{L}_{pc} = \frac{1}{M} \sum_{i=1}^{M} [ \mathcal{G}_x(s_i) - \mathcal{G}_{x}(m_i) ]^2 + [ \mathcal{G}_y(s_i) - \mathcal{G}_{y}(m_i) ]^2
\end{equation}
where $\mathcal{G}_x(\cdot)$ and $\mathcal{G}_y(\cdot)$ denote the gradient operation across horizontal and vertical axes. The reason for comparing the gradients is obvious as we do not desire to encourage the network to produce pixel intensities of $\mathcal{M}$, but the smoothness property of the pixels within each cluster. \par
\noindent \textbf{Total Variation Loss.} $\mathcal{L}_{tv}$ regularizes the model to preserve large-scale edges and textures of the image while smoothing out the noise gradients. We use the standard formulation of total variation~\cite{RUDIN1992259} in our implementation. Given the signal estimate $\mathcal{S}$, $\mathcal{L}_{tv}$ can be written as:
\begin{equation}
    \mathcal{L}_{tv} = \frac{1}{M} \sum_{i=1}^{M} | \mathcal{G}_{x}(s_i) | + | \mathcal{G}_{y}(s_i) |
\end{equation}

\section{Experiments}
\noindent \textbf{Implementation Details}. Noisy images are obtained by synthetic contamination of the clean images with additive white Gaussian noise. During the training, we randomly crop images into patches of size $64 \times 64$ and augment them by random 90$^{\circ}$ rotations and horizontal/vertical flip, however, the quantitative evaluation is reported on the full-size images of the test set. All networks are trained for 500 epochs with a batch size of 16 using Adam optimizer with default parameters. The initial learning rate is set to $10^{-4}$ and is halved every 100 epochs. For Noise2Noise and Noise2Self, we replicate the original training settings reported by the authors. We use publicly released codes to implement non-local mean (NLM) and BM3D denoising methods. It is noteworthy that all loss terms in the loss function have equal coefficients set to 1, except for the total variation term which is set to $5\times10^{-5}$. \par
\tblResults
\figQualitative

\subsection{Data}
We evaluate our proposed denoising method against state-of-the-art on three publicly available confocal laser endomicroscopy images (summarized in Table~\ref{tab:cledatasets}). Each dataset contains high-quality gray-scale images of size 1024$\times$1024 acquired by Pentax EC-3870FK imaging device during a confocal gastroscopy. Below, we provide a detail description of the datasets used in this study. \par
    \noindent \textbf{CLE100 Dataset~\cite{cle100}}. The CLE100 consists of 181 confocal fluorescence endomicroscopy images taken from 30 patients. The images have been collected from 50 different anatomical sites in the small intestine during a clinical trial to detect Celiac disease conducted at Gastroenterology and Liver Services of the Bankstown-Lidcombe Hospital (Sydney, Australia). The dataset contains images with the high-quality appearance and sharp textures. We first group images from similar anatomical sites together and then randomly split the dataset across the groups with ratio $80\%:20\%$ for training and test, respectively.\par
\noindent \textbf{CLE200 Dataset~\cite{cle200}}. The CLE200 compromises 262 images from 32 patients to diagnose Barrett’s esophagus. Images are acquired from 81 different bioptic sites at the European Oncological Institute and Veneto Institute of Oncology during clinical surveillance endoscopy of patients with the disease. All images in this dataset mostly contains dark intensity appearance with granular texture patterns. We randomly split the dataset into 212 training and 50 test images.   \par
\noindent \textbf{CLE1000 Dataset~\cite{cle1000}}. The CLE1000 is the largest publicly CLE dataset containing 1025 images from 11 patients. The dataset was collected in a recent study at Research Center of Gastroenterology and Hepatology, Romania to develop an
automatic diagnosis algorithm of colorectal cancer using fractal analysis and neural network modeling of the CLE-generated colon mucosa images. CLE1000 contains sequential video frames, hence we split the dataset across sequences not singles frames.\par

\subsection{Results}
In this section, we compare the performance of WhiteNNer against both CNN-based models and classic denoising techniques on confocal laser endomicroscopy images. Among CNN-based models, we choose Noise2Noise and Noise2Self training scheme which requires no ground truth during the training. We also train a Noise2True network to provide an estimate of the upper bound on the networks' performance in the presence of true signal. For classic methods, non-local mean (NLM) and BM3D are chosen for comparisons. \par

\figNoiseLevel

\noindent \textbf{Quantitative Evaluation.}
Table~\ref{tab:results} shows the quantitative denoising results in terms of peak signal-to-noise ratio (PSNR) and structural similarity (SSIM) for noise level $\sigma=50$. Three columns in Table~\ref{tab:results} are included to indicate whether methods exploit any additional information other than the single noisy in their algorithms. Specifically, columns denoted by $\sigma$, {\footnotesize$ \mathcal{S}$} and {\footnotesize$ \mathcal{N}'$} point out the use of noise variance, ground truth, or different realization of the noisy image, respectively.  Among the comparing methods, NLM and BM3D require the knowledge of standard deviation of the noise distribution, $\sigma$. Instead of providing the true $\sigma$ to these methods, we leverage a noise level estimator~\cite{noise_estimator} technique to predict $\hat \sigma$ from a single noisy image. \par
As can be seen in the last two rows of Table~\ref{tab:results}, our experiments reveal that WhiteNNer-2 which refers to network architecture with explicit signal and noise outputs can considerably outperform the network with only one output associated to the signal estimate, referred to as WhiteNNer-1. We relate this observation to the novel architectural regularization incorporated by allowing WhiteNNer-2 to explicitly learn two distinct latent representations for signal and noise, leading to the better reconstruction of both. We also observed that WhiteNNer-2 outperforms NLM, BM3D despite having less information about the noise realization. WhiteNNer-2 also achieves higher PSNR and SSIM than Noise2Self while neither uses any other information but the single noisy image. As one would expect, Noise2Noise and Noise2True give superior quantitative performance than WhiteNNer-2 since they both benefit from the ground truth; one for direct learning supervision and other for generating different instances of the noisy image.  \par

\noindent \textbf{Qualitative Evaluation.} 
Figure~\ref{fig:qual} visually compares the denoising results across three CLE datasets. Our experiments reveal that BM3D is highly sensitive to the estimation of the noise variance $\sigma^2$ as it produces undesired artifacts for inaccurate $\hat \sigma$ predictions. We also observe that the amount of introduced artifacts by BM3D is decreased once the true $\sigma$ is provided, however, we bound the algorithm to the estimate of $\sigma$ to make the comparisons more meaningful. Furthermore, we can observe that Noise2Self removes the noise at the cost of generating over-smoothed textures and edges around the cell bodies. In contrast, WhiteNNer-2 is able to extract the underlying signal from the noisy input with sharper edges. Comparing Noise2True and WhiteNNer-2, one can find their negligible difference intriguing as the former is trained in a fully supervised manner by considering the ground truth, as opposed to the latter which is trained only based on priors. \par

\tblStatSig
\tblAblation

\subsection{Discussions}
\noindent \textbf{Sensitivity to Noise Level}. We additionally assess the effect of noise level for all CNN-based methods on the CLE100 dataset. Figure~\ref{fig:noiselevel} depicts the average PSNR scores with respect to different noise level $\sigma \in \{15, 25,35,50,75\}$ for four different methods. It can be seen that our WhiteNNer-2 method consistently achieves higher PSNR scores across a wide range of noise levels. Also, one may notice that the performance gap between Noise2Noise/Noise2True and Noise2Self/WhiteNNer widens as the noise level increase. This observation demonstrates the impact of external information in effective denoising when the noise level is high.  \par
\noindent \textbf{Ablation Study}. We examine the effect of loss terms in efficient denoising on CLE100 with $\sigma=50$. To do so, we train each model configuration for 100 epochs and compare their performance on a held-out test data. In Table~\ref{tab:ablation}, it can be seen that each prior plays an efficient role in pushing the network to produce signals of higher quality. In particular, we note that the PSNR score improves by 2.74dB once the auto-correlation term is added to the loss function. Furthermore, including the stationary prior term encourages the network to generate spatially consistent noise components which consequently leads to better estimate of the signal. The best performance is achieved when all prior term for signal and noise are included in the loss function.\par

\noindent \textbf{Statistical Significance}.We perform paired t-test to compare our approach with NLM, BM3D and Noise2Self for $\sigma=50$ on CLE100. As shown in Table~\ref{tab:stats}, the $p$-values are less than 0.05 for all experiments, which means that WhiteNNer is capable of producing results that are significantly better than those of comparing methods.
\section{Conclusion}
In this paper, we propose a novel CNN-based model to address the problem of blind image denoising via joint exploitation of signal and noise priors. We propose a novel loss function that encourages the model outputs to conform to signal and noise constraints. For signal, we leverage piecewise constancy and total variation information priors. Additionally, we design auto-correlation and stationary regularization terms to model the noise whiteness prior. Given the accurate and discriminative priors, our network is the first model that is capable of decoupling the signal and noise components without ground truth information. Our experimental results show superiority of our proposed model against state-of-the-art methods in terms of PSNR and SSIM on three publicly available confocal laser endomicroscopy datasets.

\noindent\textbf{Acknowledgments.} Thanks to the NVIDIA Corporation for the donation of Titan X GPUs used in this research, the Collaborative Health Research Projects (CHRP) for funding, and ComputeCanada for computational resources.

{\small
\bibliographystyle{ieee}
\bibliography{egbib}
}

	{\small
		\bibliographystyle{ieee}
		\bibliography{bib}
	}
	
\end{document}